\let\TeXyear\year
\documentclass{IEEEaccess}

\let\year\TeXyear
\usepackage{cite}
\usepackage{amsmath,amssymb,amsfonts}
\usepackage{graphicx}
\usepackage{textcomp}
\def\BibTeX{{\rm B\kern-.05em{\sc i\kern-.025em b}\kern-.08em
    T\kern-.1667em\lower.7ex\hbox{E}\kern-.125emX}}
\history{Date of publication xxxx 00, 0000, date of current version xxxx 00, 0000.}
\doi{10.1109/ACCESS.2017.DOI}

\usepackage{textcomp}
\usepackage{xcolor}

\usepackage{amsfonts}
\usepackage{url}
\usepackage{tikz}

\usetikzlibrary{shapes.geometric, arrows, positioning, backgrounds, fit, shadows, decorations.pathreplacing, arrows.meta}
\tikzset{
    startstop/.style={rectangle, rounded corners=4pt, minimum width=3cm, minimum height=1cm, text centered, draw=black!80, fill=blue!20, thick},
    process/.style={rectangle, rounded corners=4pt, minimum width=3cm, minimum height=1cm, text centered, text width=5cm, draw=black!80, fill=blue!10, thick},
    decision/.style={diamond, minimum width=3cm, minimum height=1cm, text centered, draw=black!80, fill=orange!15, thick, aspect=2},
    arrow/.style={thick,->,>=latex, shorten >=2pt, shorten <=2pt}
}
\usepackage[edges]{forest}

\usepackage[utf8]{inputenc}
\usepackage{booktabs}
\usepackage{hyperref}
\usepackage{mathtools}
\usepackage{pgfplots}
\pgfplotsset{compat=1.18, width=0.85\columnwidth, height=0.45\columnwidth, grid=major, grid style={dashed,gray!30}, legend style={font=\footnotesize,fill=white,draw=black!50}, tick label style={font=\footnotesize}, label style={font=\small}}
\usepackage{caption}
\usepackage{algorithm}
\usepackage{algpseudocode}
\IEEEoverridecommandlockouts
\begin{document}
\NewSpotColorSpace{PANTONE}
\AddSpotColor{PANTONE}{PANTONE3015C}{PANTONE\SpotSpace 3015\SpotSpace C}{1 0.3 0 0.2}
\SetPageColorSpace{PANTONE}
\title{Novel Iterative Construction Methods for the Blocking Job Shop Scheduling Problem}

\author{\uppercase{Adel Dabah}\authorrefmark{1}, 
\uppercase{Karima Rihane}\authorrefmark{3}, 
\uppercase{Hocine Saadi}\authorrefmark{2}, 
\uppercase{Andreas Herten}\authorrefmark{1}, 
\uppercase{Farouk Benslimane}\authorrefmark{3}, 
\uppercase{Mohammed Lamine Bahmani}\authorrefmark{3}, 
and \uppercase{Abdelhakim AitZai}\authorrefmark{2}}

\address[1]{J\"ulich Supercomputing Centre (JSC), Forschungszentrum J\"ulich, 52425 J\"ulich, Germany (e-mail: \{a.dabah, a.herten\}@fz-juelich.de)}
\address[2]{University of Sciences and Technology Houari Boumediene (USTHB), Algiers, Algeria (e-mail: \{hsaadi2, a.aitzai\}@usthb.dz)}
\address[3]{University of Sciences and Technology Houari Boumediene (USTHB), Algiers, Algeria (e-mail: \{k.rihane, farouk.benslimane, mohammedlamine.bahmani\}@etu.usthb.dz)}


\markboth
{Dabah \headeretal: Novel Iterative Construction Methods for the Blocking Job Shop Scheduling Problem}
{Dabah \headeretal: Novel Iterative Construction Methods for the Blocking Job Shop Scheduling Problem}

\corresp{Corresponding author: Adel Dabah (e-mail: a.dabah@fz-juelich.de).}

\begin{abstract}
The Blocking Job-Shop Scheduling Problem (BJSSP) arises in modern and complex manufacturing, production, logistics, and service where no intermediate storage is allowed between consecutive operations. This creates a significant challenge for meta-heuristics due to the low ratio of feasible to explored solutions when solving the problem. To address this problem efficiently, we propose three new beam-search-based heuristics: the \emph{Beam Search Iterative Construction Heuristic} (BS-ICH), its CPU-parallel extension \emph{Parallel Multi-Strategy Beam Search} (PMS-BS), and a GPU-accelerated variants \emph{G-PMS-BS}. BS-ICH constructs feasible schedules by iteratively extending partial solutions, while maintaining a beam of width $k$ to preserve multiple high-quality partial schedules. PMS-BS runs hundreds of parallel BS-ICH instances with machine-biased diversity to expand the search space and escape local optima. G-PMS-BS offloads the beam expansion onto massively parallel GPU hardware using a two-phase kernel architecture that separates lightweight scoring from targeted state reconstruction, enabling scaling to instances with 2{,}000 operations. A hybrid CPU+GPU mode further exploits idle host cores for concurrent exploration, using load-balancing strategy to minimize synchronization overhead. G-PMS-BS achieves a $44\times$ speedup over the CPU baseline. Experiments on all standard Lawrence and Taillard instances demonstrate that G-PMS-BS establishes new best-known results for 22 Lawrence benchmarks and 77 Taillard instances, with makespan reductions of up to 13\% on the largest 100$\times$20 instances.
\end{abstract}
\maketitle
\section{Introduction}

The Job‐Shop Scheduling Problem (JSSP) is a well studied combinatorial optimization with wide‐ranging applications in manufacturing, logistics, and service systems. In its classical form, each job traverses its own sequence of machines with unlimited intermediate storage. However, in many real‐world settings such as semiconductor fabrication, chemical batch processing, and automated assembly lines, space prohibits any storage between successive operations. This zero‐buffer variant, known as the Blocking Job‐Shop Scheduling Problem (BJSSP), introduces additional coupling between machines: a job that completes on a given machine blocks it until its next machine is free. The resulting dependencies exacerbate the combinatorial explosion of the search space and invalidate many heuristic rules designed for finding solutions to the classical JSSP.

Over the past decades, exact methods~\cite{d1,d2,d3,dabah2018hybrid} and metaheuristics~\cite{dabah2019efficient,d4,10} have been proposed to solve BJSSP. While exact solvers can optimally solve small instances, their runtime scales poorly with problem size. Metaheuristics offer better scalability, but often suffer from excessive infeasible solution exploration and high-cost parameter tuning. In addition, feasibility recovery methods are time-consuming and prevent the efficient exploration of the search space. Moreover, hybrid and learning‐based methods such as~\cite{rihane2024adapted,rihane2025learning,9870384} have recently shown promising results but at the cost of increased algorithmic complexity due to the exponential increase in states/actions. 

In this paper, we propose three complementary beam‐search construction heuristics that bridge the gap between purely greedy dispatching rules and full enumeration. First, the K‐Best Iterative Construction Heuristic (BS-ICH) maintains a beam of the \(K\) most promising partial schedules, extending each by all feasible next operations under blocking constraints and pruning back to the top \(K\) by partial makespan. Second, to enhance exploration and prevent convergence to similar beams, we introduce \emph{Parallel Multi‐Strategy Beam Search} (PMS‐BS), which runs \(L\) independent BS-ICH instances in parallel. Each instance incorporates a distinct machine‐priority bias and a small random injection when selecting its top‐\(K\) candidates, thereby diversifying the search across different regions of the solution space. 
The novelty of BS-ICH lies in enforcing feasibility/exploration balance throughout the beam expansion, avoiding recovery phases of metaheuristics. PMS-BS adds a parallel diversification with machine-priority biases to explicitly integrate feasibility, parallelism, and diversification. 

To further scale the proposed approach to very large instances, we introduce a GPU-accelerated variant (G-PMS-BS) that offloads the beam expansion onto massively parallel GPU hardware. The GPU implementation uses a two-phase kernel architecture that separates lightweight scoring from targeted state reconstruction, enabling scaling to instances with 2,000 operations. A hybrid CPU+GPU mode further exploits otherwise-idle host cores with a load-balancing mechanism to minimize synchronization overhead.

The main contributions of this work are: 
\begin{itemize}
    \item  Novel construction-based heuristic (BS-ICH) which directly constructs feasible schedules by maintaining the k-best partial schedule at each step of the construction.   
    \item Parallel multi-strategy framework (PMS-BS), which runs multiple beam search instances in parallel to enhance exploration, escape local optimal that traps BS-ICH, and improve solution robustness. 
    \item GPU-accelerated variants (G-PMS-BS) that offload beam expansion onto massively parallel GPU hardware. A two-phase kernel architecture to overcome GPU memory constraints, and a hybrid CPU+GPU version uses otherwise-idle CPU cores for additional exploration with load-balancing mechanism.  
    \item Experiments on all 40~Lawrence and 80~Taillard benchmarks with new best results for large number of instances.
    \item GPU-accelerated variant (G-PMS-BS) that offloads beam expansion onto NVIDIA A100 GPUs with a two-phase kernel architecture, achieving a 44× speedup over the CPU baseline.
\end{itemize}

The remainder of this paper is organized as follows. 
Section \ref{sec:BJSSP} introduces the Blocking Job Shop Scheduling Problem (BJSSP) and related works. Section \ref{sec:beam} presents the proposed Beam Search Iterative Construction Heuristic (BS-ICH). 
Section \ref{sec:pms-bs} extends this approach to the Parallel Multi-Strategy Beam Search (PMS-BS). 
Section \ref{sec:gpu} describes the GPU-accelerated version (G-PMS-BS), including the two-phase pipeline and hybrid execution strategy. describes the GPU-accelerated versions.
Section \ref{sec:experiments} reports the experimental evaluation on both Lawrence and Taillard benchmarks and compares our results with state-of-the-art methods. 
Finally, Section \ref{sec:conclusion} concludes the paper and outlines future research directions.

\section{Blocking Job Shop Scheduling Problem}
\label{sec:BJSSP}

The classical \emph{Job-Shop Scheduling Problem} (JSSP) considers a set of jobs \(\mathcal{J} = \{1, \dots, J\}\), each composed of an ordered sequence of operations to be executed on a set of machines \(\mathcal{M} = \{1, \dots, M\}\). Let \(\mathcal{O} = \{o_1, o_2, \dots, o_{n \times m}\}\) denote the complete set of operations, where each operation \(o_i\) is assigned to a machine \(M(i) \in \mathcal{M}\) and requires a non-preemptive processing time \(p_i > 0\). For each job, the operations follow a fixed routing, inducing precedence constraints between successive operations of that job. 
In the classical JSSP, it is assumed that intermediate buffers between machines are of unlimited capacity, allowing operations to complete and wait without restriction. This assumption, however, is often impractical. The \emph{Blocking Job-Shop Scheduling Problem} (BJSSP) removes this assumption by enforcing a \emph{zero-buffer} constraint: an operation must remain on its current machine (thus blocking it) until its successor is able to begin on the next machine.

Let \(\sigma(i)\) denote the index of the successor operation of \(o_i\) in the same job (if it exists). A \emph{schedule} assigns a start time \(t_i \ge 0\) to each operation \(o_i \in \mathcal{O}\), with completion time \(c_i \ge t_i + p_i\), such that the following constraints are satisfied:

\begin{enumerate}
  \item \textbf{Precedence constraints:}  
  Each operation must start after its predecessor finishes:
  \[
    t_{\sigma(i)} \ge t_i + p_i 
    \quad \forall\, o_i \in \mathcal{O} \text{ with } \sigma(i) \text{ defined}.
  \]

  \item \textbf{Machine capacity (no overlap):}  
  No two operations may overlap on the same machine. For any pair \(o_i, o_j\) with \(M(i) = M(j)\) and \(i \ne j\), we impose:
  \[
    \text{Either } t_i \ge t_j + p_j \quad \text{or} \quad t_j \ge t_i + p_i.
  \]

  \item \textbf{Blocking constraint (zero-buffer):}  
  A job cannot leave its current machine until its next operation is ready to start. Thus, for \(o_i\) with defined \(\sigma(i)\), we require: 
  \[
    c_i = t_{\sigma(i)}.
  \]  
  If \(\sigma(i)\) is undefined (i.e., \(o_i\) is the last operation of its job), then its completion time is: 
  \[
    c_i = t_i + p_i.
  \]
\end{enumerate}

The objective is to minimize the makespan, i.e., the time when the last operation finishes:
\[
  C_{\max} = \max_{o_i \in \mathcal{O}}\, (c_i).
\]

The BJSSP can also be modeled using an alternative graph representation introduced by Mascis and Pacciarelli~\cite{1}, which generalizes the disjunctive graph model of Roy and Sussman~\cite{4}. In this model, the problem is represented as a directed graph \(G = (N, F, A)\), where:
\begin{itemize}
  \item \(N = \mathcal{O} \cup \{\text{start}, \text{end}\}\) is the node set including all operations plus two dummy nodes.
  \item \(F\) is the set of fixed arcs representing precedence constraints between successive operations of a job, where each arc \((o_i, o_j)\) has weight \(f_{ij} = p_i\).
  \item \(A\) is a set of alternative pairs representing the processing order between concurrent operations.
\end{itemize}

The blocking constraint introduces implicit dependencies between operations across machines, as a delay in starting a successor operation forces the predecessor to remain on the machine, potentially preventing other jobs from accessing it. This inter-machine coupling makes BJSSP significantly more difficult than its classical counterpart.

In the BJSSP, a swap situation represents a deadlock involving two or more operations forming a zero-length cycle in the disjunctive graph representation. The BJSSP has two variants: Blocking with No Swap (BNS), where solutions with swap are considered infeasible. In contrast, the Blocking with Swap (BWS) variant allows all operations involved in a cycle to move simultaneously to their respective next machines. In this paper, we consider the BWS variant which is more challenging since it introduces additional inter-machine dependencies.

\subsection{Related Works}

Hall and Sriskandarajah~\cite{2} presented a survey on machine scheduling problems involving blocking and no-wait constraints. Mati \textit{et al.}~\cite{3} proposed a tabu search method for the BJSS, applying classical permutation on the critical path. Their approach yielded relative errors between 3\% and 9\% compared to the optimal solutions for benchmark instances with ten jobs and ten machines (10$\times$10).

Mascis and Pacciarelli~\cite{1} studied job shop problems with blocking and no-wait constraints using an alternative graph model. Based on this model, they developed four constructive heuristics and provided results for 58 benchmark instances. They also optimally solved all 10$\times$10 instances using a branch-and-bound method built upon their heuristics.

Meloni \textit{et al.}~\cite{5} introduced a Rollout metaheuristic based on the alternative graph model in~\cite{1} to solve classical job shop, blocking job shop, and no-wait job shop problems. Pham and Klinkert~\cite{7} addressed hospital resource scheduling by modeling the problem as a Multi-Mode Blocking Job Shop.

Groflin and Klinkert~\cite{8} proposed a tabu search neighborhood structure based on reversing an alternative arc in the critical path. They later extended their approach to the general job shop problem by introducing takeover and handover times for each operation~\cite{groflin2011flexible}. Oddi \textit{et al.}~\cite{9} applied an Iterative Flattening Search method to the BJSS, achieving state-of-the-art results and surpassing those of Groflin and Klinkert~\cite{8}.

AitZai \textit{et al.}~\cite{10} presented both a Branch-and-Bound method based on graph theory and a genetic algorithm with binary encoding to solve small-scale BJSSP benchmark instances. Mati and Xie~\cite{14} proposed a tabu search algorithm combined with a geometric approach for the BJSSP with no-swap constraints. Pranzo and Pacciarelli~\cite{6} developed an Iterated Greedy (IG) metaheuristic for the BJSS, improving most of the results reported in previous works.

Dabah \textit{et al.}~\cite{d3} proposed a parallel Branch-and-Bound algorithm leveraging cluster-based computing. The approach functions as both an exact and approximate method, reporting optimal solutions for ten previously unsolved benchmark instances and improving results for 22 others. In~\cite{d1,d2,dabah2018hybrid}, the authors introduced several parallelization strategies that harness both multi-core CPU and GPU architectures, achieving speedups of up to 100×. In~\cite{d4}, they proposed a neighborhood function tailored to the blocking constraint based on a reconstruction strategy. A parallel version of this approach with multiple search strategies is proposed in~\cite{dabah2019efficient}, where the authors show that running multiple tabu search instances with different machine-priority biases and exchanging the best makespan significantly improves solution quality. This multi‑strategy paradigm directly inspired the PMS-BS and G-PMS-BS methods presented in this paper.

Recent advances in machine learning have introduced novel approaches to tackle the BJSSP. Rihane \textit{et al.}~\cite{rihane2025learning} provided a comprehensive survey of learning-based methods for the classical JSSP. Rihane \textit{et al.}~\cite{9870384} proposed a learning-based selection process for B\&B, training a model on small and medium BJSSP instances to automatically derive branching rules. Rihane \textit{et al.}~\cite{rihane2024adapted} developed the first reinforcement learning framework for BJSSP, modeling the problem as a Markov decision process and employing Q-Learning with tailored reward functions.

\section{Beam Search Iterative Construction Heuristic}
\label{sec:beam}

In the following, we introduce a new concept and category of Iterative Construction Heuristic (ICH) that can be applied for all optimization problems. In the following we will see its application for the Blocking Job-Shop Scheduling Problem (BJSSP). 

The proposed method has its roots in our earlier work on large‑scale MIMO detection, where the sphere decoder on FPGAs was accelerated by maintaining a limited set of promising candidates at each layer of the search tree~\cite{hassan2023signal, dabah2023efficient}. We later extended this idea to the exact graph edit distance (GED) problem, showing that a K‑best construction strategy not only improves accuracy but also scales efficiently for large graphs~\cite{dabah2021efficient}. 

When we first attempted to apply a similar tree‑based approach (breadth‑first) to the BJSSP, we observed that most of the nodes kept in a tree level after pruning eventually led to infeasible schedules because of the blocking constraints; the method was therefore impractical. To keep the K‑best philosophy, we instead replaced the tree based model with a construction‑based approach that never generates infeasible solutions. 

At the core of the method, we have a construction-based heuristic that extends one partial schedule with one operation from a set of possible next-operations that can be executed on machines, taking into consideration all BJSSP constraints. You can see the method like priority rule based construction.  Therefore, we extend a partial schedule with one operation at each iteration until reaching a complete solution. Note that all schedules generated by ICH are feasible, which was the main reason for choosing it as the core method.


\begin{algorithm}[h!]
\caption{\small Beam Search Iterative Construction Heuristic}
\label{alg:beam}
\scalebox{0.9}{
\begin{minipage}{11.5cm}
\begin{algorithmic}[1]
\Require Jobs \(\mathcal{J}\), machines \(\mathcal{M}\), processing times \(\{p_i\}\),
beam width \(K\)
\State Initialize beam: \(\mathcal{B}_0 \gets \{\varnothing\}\)
\For{\(t = 0\) to \( |\mathcal{O}| - 1\)}
\State select machine $m$
  \State \(\mathcal{Q} \gets \emptyset\)
  \ForAll{\(Ps \in \mathcal{B}_t\)}
    \ForAll{\(o \in \mathrm{NextOps}(Ps),\ where\ M(o) = m\)}
      \State \(Ps' \gets Ps \cup \{o\}\)
      \State Evaluate \(\Phi(Ps') \gets C_{\max}(Ps')\)
      \State \(\mathcal{Q} \gets \mathcal{Q} \cup \{Ps'\}\)
    \EndFor
  \EndFor
  \State \(\mathcal{B}_{t+1} \gets\) top(K) \textit{Ps} in \(\mathcal{Q}\) by lowest \(\Phi\)
\EndFor
\State \Return \(\arg\min_{Ps \in \mathcal{B}_{|\mathcal{O}|} }C_{\max}(Ps)\)
\end{algorithmic}
\end{minipage}
}
\end{algorithm}

On top of ICH, we add a beam-search level, where we maintain at each step of the algorithm, the \(K\) most promising partial schedules instead of one. At each iteration, we first select a machine $m$ and then we extend all \(K\) partial schedules with all their possible next-operations on machine $m$, creating thousands of feasible partial schedules. Selecting a single machine is crucial to avoid exploring duplicate solutions. After that, we evaluate all resulted partial-schedules and prune back to the best \(K\) based on the makespan. We repeat this process until we have a full $K$ schedules (solution to our problem), where we return the best solution in terms of makespan. One possible improvement to this approach would be to use a heuristic to estimate the remaining cost, which would result in a more accurate selection.

Formally, we consider \(\mathcal{B}_t \subseteq \mathbb{S}\): beam at iteration \(t\), with \(|\mathcal{B}_t| \le K\).
\begin{itemize}
  \item \emph{Partial schedule} \(Ps\) assigns start/end times to a subset of operations \(\mathrm{ops}(Ps)\).
  \item \(\mathrm{NextOps}(Ps)\): operations whose predecessors are scheduled and whose machines are available for processing.
  \item $C_{\max}(Ps)$: the makespan of partial schedule $Ps$.
\end{itemize}

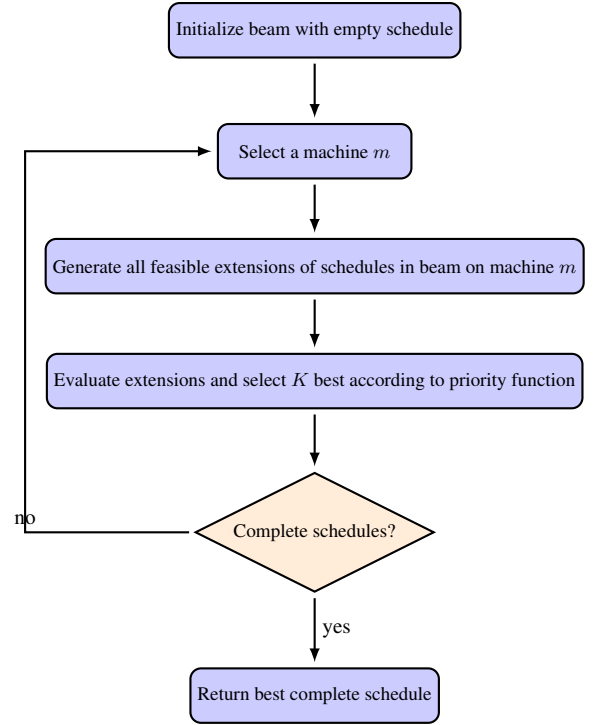
\begin{figure}[h!]
    \centering
    \begin{tikzpicture}[node distance=1.50cm, auto, >=latex, scale=0.8, transform shape]
        \tikzstyle{block} = [rectangle, rounded corners=4pt, minimum width=3.2cm, minimum height=0.9cm, text centered, draw=black, fill=blue!20, thick, font=\small];
        \tikzstyle{decision} = [diamond, aspect=2, minimum width=2.5cm, minimum height=0.8cm, text centered, draw=black, fill=orange!15, thick, font=\small];
        \tikzstyle{arrow} = [thick,->,>=latex, shorten >=2pt, shorten <=2pt];
        
        \node (start) [block] {Initialize beam with empty schedule};
        \node (select_machine) [block, below of=start, yshift=-0.5cm] {Select a machine $m$};
        \node (gen_ext) [block, below of=select_machine, yshift=-0.4cm] {Generate all feasible extensions of schedules in beam on machine $m$};
        \node (evaluate) [block, below of=gen_ext, yshift=-0.4cm] {Evaluate extensions and select $K$ best according to priority function};
        \node (dec1) [decision, below of=evaluate, yshift=-1.0cm] {Complete schedules?};
        \node (return) [block, below of=dec1, yshift=-1.2cm] {Return best complete schedule};
        
        \draw [arrow] (start) -- (select_machine);
        \draw [arrow] (select_machine) -- (gen_ext);
        \draw [arrow] (gen_ext) -- (evaluate);
        \draw [arrow] (evaluate) -- (dec1);
        \draw [arrow] (dec1) -- node[anchor=west, midway] {yes} (return);
        \draw [arrow] (dec1.west) -- ++(-2.8,0) node[anchor=south] {no} |- (select_machine.west);
    \end{tikzpicture}
    \caption{Beam Search Iterative Construction Heuristic with machine selection.}
    \label{fig:beam_search_flowchart}
\end{figure}

Algorithm~\ref{alg:beam} and Figure~\ref{fig:beam_search_flowchart} show the steps of the proposed method. At each iteration, up to \(K\) partial schedules are extended by up to \(J\) operations. The complexity in worst case is $O(|\mathcal{O}| \cdot K \cdot J)$ assuming constant-time $C_{\max}$ calculation and sorting. Our approach balances exploration and runtime by adjusting the value of parameter $K$. 
\begin{itemize}
  \item \(K = 1\): recovers greedy rule-based construction.
  \item \(K \in [5, 20]\): empirically yields \(5\%-10\%\) better makespans with moderate overhead.
\end{itemize}
Therefore, our approach offers a tunable trade-off between greedy construction and exhaustive search, making it a scalable and effective baseline for solving BJSSP.

In the following we give an illustrative example in Table~\ref{tab:example} with three jobs and two machines .
\begin{table}[h!]
\centering
\scalebox{0.9}{
\begin{tabular}{@{}ll@{}}
\toprule
\textbf{Job} & \textbf{Operation Sequence (Machine, Duration)} \\
\midrule
\( J_1 \) & \( (M_1, 3) \rightarrow (M_2, 2) \) \\
\( J_2 \) & \( (M_2, 2) \rightarrow (M_1, 4) \) \\
\( J_3 \) & \( (M_1, 2) \rightarrow (M_2, 3) \) \\
\bottomrule
\end{tabular}
}
\caption{Job operation sequences for the BJSSP example}
\label{tab:example}
\end{table}

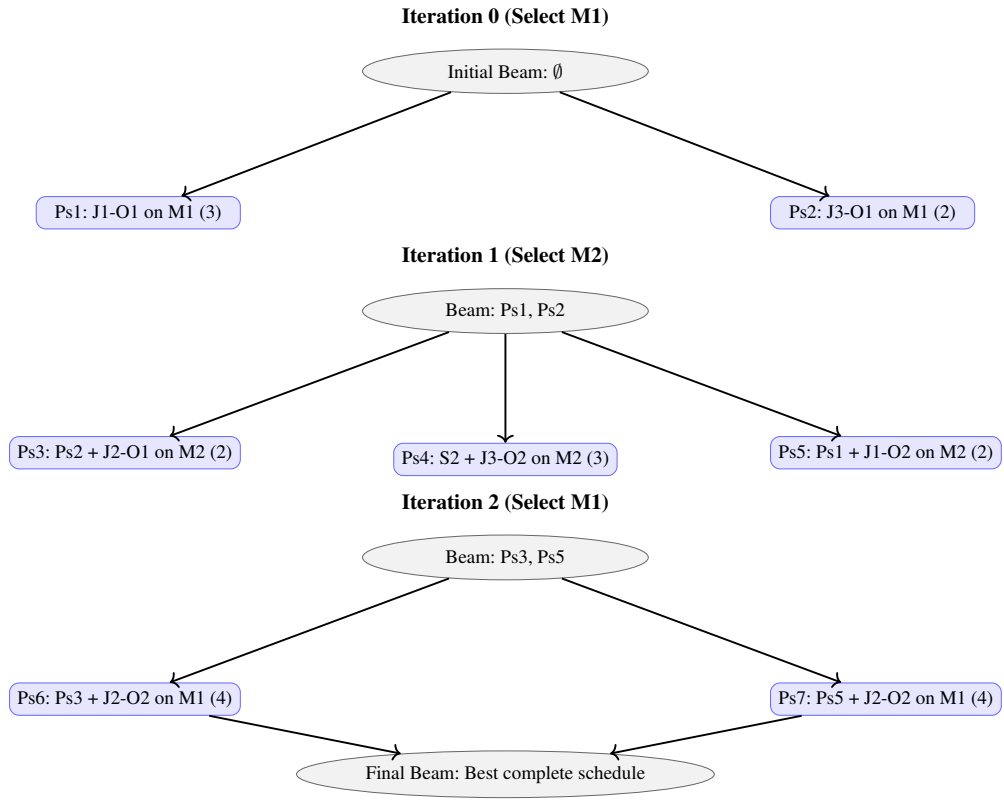
\begin{figure*}[h!]
\centering
\centering
\scalebox{0.9}{
\begin{tikzpicture}[
  node distance=1.6cm and 1.8cm,
  op/.style={rectangle, draw=blue!60, fill=blue!10, rounded corners, minimum width=3cm, align=center, font=\footnotesize},
  beam/.style={ellipse, draw=black!60, fill=gray!10, minimum width=4.2cm, align=center, font=\footnotesize},fill=gray!10,
  label/.style={font=\bfseries\small},
  arrow/.style={->, thick}
]

\node[label] (i0label) at (0,0) {Iteration 0 (Select M1)};
\node[beam] (b0) [below=0.2cm of i0label] {Initial Beam: $\emptyset$};
\node[op] (s1) [below left=of b0, xshift=-0.6cm] {Ps1: J1-O1 on M1 (3)};
\node[op] (s2) [below right=of b0, xshift=0.6cm] {Ps2: J3-O1 on M1 (2)};
\draw[arrow] (b0) -- (s1);
\draw[arrow] (b0) -- (s2);

\node[label] (i1label) [below=3.0cm of i0label] {Iteration 1 (Select M2)};
\node[beam] (b1) [below=0.2cm of i1label] {Beam: Ps1, Ps2};
\node[op] (s3) [below left=of b1, xshift=-0.6cm] {Ps3: Ps2 + J2-O1 on M2 (2)};
\node[op] (s4) [below=of b1] {Ps4: S2 + J3-O2 on M2 (3)};
\node[op] (s5) [below right=of b1, xshift=0.6cm] {Ps5: Ps1 + J1-O2 on M2 (2)};
\draw[arrow] (b1) -- (s3);
\draw[arrow] (b1) -- (s4);
\draw[arrow] (b1) -- (s5);

\node[label] (i2label) [below=3.1cm of i1label] {Iteration 2 (Select M1)};
\node[beam] (b2) [below=0.2cm of i2label] {Beam: Ps3, Ps5};
\node[op] (s6) [below left=of b2, xshift=-0.6cm] {Ps6: Ps3 + J2-O2 on M1 (4)};
\node[op] (s7) [below right=of b2, xshift=0.6cm] {Ps7: Ps5 + J2-O2 on M1 (4)};
\draw[arrow] (b2) -- (s6);
\draw[arrow] (b2) -- (s7);

\node[beam] (bfinal) [below=2.5cm of b2] {Final Beam: Best complete schedule};

\draw[arrow] (s6) -- (bfinal);
\draw[arrow] (s7) -- (bfinal);

\end{tikzpicture}
}
\caption{K-Best Iterative Construction Heuristic (K=2) — Beam Search with Alternating Machine Selection}
\label{fig:beamsearch}
\end{figure*}

\paragraph{Complexity Analysis}

As the foundational algorithmic core of our proposed methods, the complexity of BS-ICH defines the scalability ceiling for all subsequent parallel CPU and GPU-accelerated extensions. Unlike exhaustive search methods, BS-ICH achieves a polynomial runtime by exploiting the construction-based feasible expansion and restricting the search to a fixed beam width \(K\).

Let \(Nop = |\mathcal{O}|\) denote the total number of operations, \(M = |\mathcal{M}|\) the number of machines, and \(J = |\mathcal{J}|\) the number of jobs. The algorithm iterates exactly \(Nop\) times, once for each scheduled operation. At iteration \(t\), the beam \(\mathcal{B}_t\) contains at most \(K\) partial schedules. For each partial schedule, the algorithm considers only the feasible next operations that belong to the single selected machine \(m\). In the worst case, this may yield at most \(J\) candidate extensions per partial schedule. Consequently, the number of child schedules generated per iteration is bounded by \(K \cdot J\).

For each generated child schedule, we update the partial makespan \(C_{\max}\). Since the schedule is constructed incrementally, this update is performed in \(O(1)\) time by maintaining the machine release times and job completion times. After all children are generated, the algorithm selects the \(K\) best candidates. Using a Radix Sort, this top-\(K\) operation requires \(O(K \cdot J \log K)\) time per iteration. Therefore, the total worst-case time complexity of BS-ICH is:

\[
O\!\left( N \cdot K \cdot J \log K \right).
\]

When \(K\) is treated as a small constant relative to the problem size (e.g., \(K \leq 10,000\) in our experiments), the logarithmic factor is negligible, and the complexity simplifies to the bound stated in Algorithm~1: \(O(N \cdot K \cdot J)\). 

The space complexity is dominated by the storage of the beam. Storing \(K\) partial schedules, each containing up to \(N\) operations with their associated timing metadata, requires \(O(N \cdot K)\) memory. The candidate queue \(\mathcal{Q}\) temporarily holds at most \(K \cdot J\) children, adding another \(O(K \cdot J)\) term, which is bounded by \(O(N \cdot K)\) since \(J \leq N\).

This polynomial complexity is the direct consequence of our two key design choices: (i) the ICH construction rule that guarantees feasibility without backtracking, and (ii) the machine-serialized expansion that prevents duplicate exploration of identical partial schedules. This core complexity remains unchanged when we extend BS-ICH to PMS-BS and G-PMS-BS. Thus, the polynomial efficiency of BS-ICH provides the essential foundation that allows the parallel and GPU-accelerated variants to scale to instances with 2,000 operations.

Figure~\ref{fig:beamsearch} illustrates the Beam Search Iterative Construction Heuristic for a Blocking Job-Shop Scheduling Problem with three jobs, two machines, and a beam width $K = 2$. At each iteration, a single machine is selected, and the algorithm expands the current beam of partial schedules by inserting only feasible operations assigned to that machine. The figure shows the iterative growth of the solution space: starting from an empty schedule, the beam is first expanded using operations on machine $M_1$, producing two partial schedules. In the next iteration, machine $M_2$ is selected, and each beam candidate is extended with operations executable on $M_2$, subject to precedence and blocking constraints. This process continues by alternating machines and pruning to retain only the top $K$ partial schedules based on a priority metric such as makespan. The final beam contains the most promising complete schedules, from which the best is selected. This figure emphasizes the guided and constrained expansion mechanism central to the beam search strategy.

A limitation of BS-ICH is its machine selection, leading the algorithm to converge to a suboptimal schedule. This motivates our parallel PMS-BS extension, which explicitly introduces diversity strategies.

\section{Parallel Multi-Strategy Beam Search (PMS-BS)}
\label{sec:pms-bs}

To improve solution diversity and avoid local optima, we propose the \emph{Parallel Multi-Strategy Beam Search (PMS-BS)}. Indeed, selecting a machine $m$ at each iteration helps avoid duplicate solution exploration. However, it also reduces the number of explored branches with many potentially good solutions pruned early. As a result, the search may converge to a local optimum. To solve this issue, PMS-BS launches $L$ fully independent BS-ICH instances, each guided by a distinct machine-priority bias and a controlled amount of randomness. Although the instances explore the search space independently, they exchange the best makespan found so far, allowing each to prune partial schedules whose estimated makespan exceeds the current global best, thus reducing non-promising exploration of suboptimal solutions. 

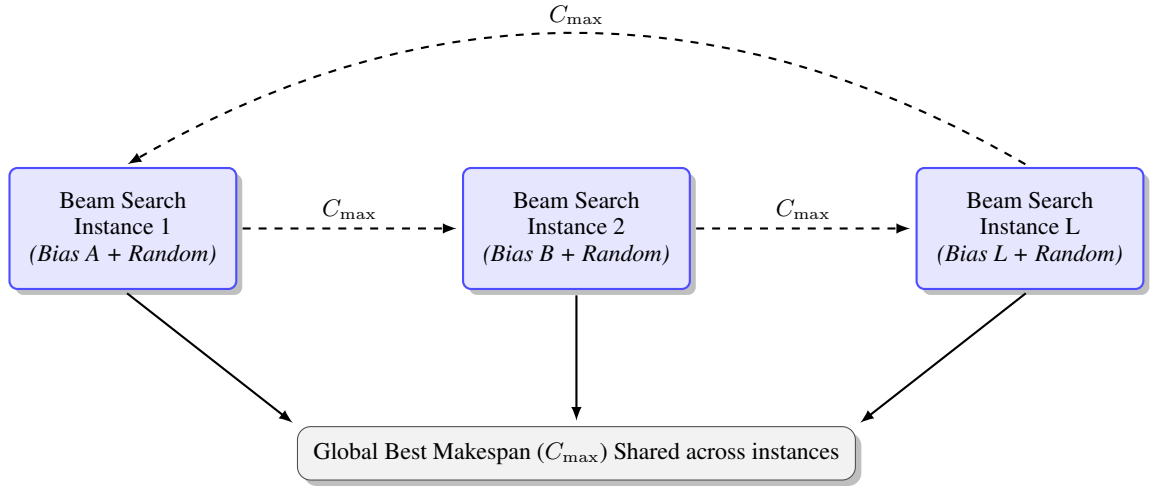
\begin{figure*}[t!]
\centering
\begin{tikzpicture}[
    instance/.style={rectangle, draw=blue!70, fill=blue!10, thick, minimum width=3.0cm, minimum height=1.6cm, rounded corners=3pt, drop shadow},
    arrow/.style={thick,->,>=latex, shorten >=2pt, shorten <=2pt},
    textnode/.style={align=center, font=\small},
    box/.style={draw=black!70, rounded corners=5pt, fill=gray!10, inner sep=6pt, drop shadow},
    every node/.style={font=\small}
]
\node[instance] (inst1) at (0,0) {};
\node[instance] (inst2) at (6,0) {};
\node[instance] (inst3) at (12,0) {};
\node[textnode] at (0,0) {Beam Search\\ Instance 1\\ \textit{(Bias A + Random)}};
\node[textnode] at (6,0) {Beam Search\\ Instance 2\\ \textit{(Bias B + Random)}};
\node[textnode] at (12,0) {Beam Search\\ Instance L\\ \textit{(Bias L + Random)}};
\draw[arrow, dashed] (inst1.east) -- (inst2.west) node[midway, above] {$C_{\max}$};
\draw[arrow, dashed] (inst2.east) -- (inst3.west) node[midway, above] {$C_{\max}$};
\draw[arrow, dashed, bend right=30] (inst3.north) to node[midway, above] {$C_{\max}$} (inst1.north);
\node[box, below=1.8cm of inst2] (global) {Global Best Makespan ($C_{\max}$) Shared across instances};
\draw[arrow] (inst1.south) -- (global.north west);
\draw[arrow] (inst2.south) -- (global.north);
\draw[arrow] (inst3.south) -- (global.north east);
\end{tikzpicture}
\caption{Parallel Multi-Strategy Beam Search (PMS-BS). Each instance performs an independent beam search using a unique machine-priority bias and randomness. They periodically exchange the best known makespan $C_{\max}$ to prune unpromising schedules.}
\label{fig:pmsbs}
\end{figure*}

An overview of the proposed method is provided in Figure~\ref{fig:pmsbs} and detailed in Algorithm~\ref{alg:pms-bs}.

Let \(L\) be the number of parallel instances.
\(\mathcal{B}_t^{(\ell)}\) beam at iteration \(t\) of instance \(\ell\), for \(\ell = 1,\dots,L\).
\(\pi^{(\ell)}: \mathcal{M} \to \mathbb{R}_+\) machine-priority weights for instance \(\ell\).
\(\epsilon\) a probability of random injection during beam pruning.

\begin{algorithm}[h!]
\caption{\small PMS-BS: Parallel Multi-Strategy Beam Search}
\label{alg:pms-bs}
\scalebox{0.7}{
\begin{minipage}{11.5cm}
\begin{algorithmic}[1]
\Require Jobs \(\mathcal{J}\), machines \(\mathcal{M}\), processing times \(\{p_i\}\), beam width \(K\), number of strategies \(L\), diversification probability \(\epsilon\), strategy biases \(\{\pi^{(\ell)}\}_{\ell=1}^L\)
\For{each \(\ell = 1\) to \(L\) \textbf{in parallel}}
  \State Initialize beam: \(\mathcal{B}_0^{(\ell)} \gets \{\varnothing\}\)
\EndFor
\For{\(t = 0\) to \( |\mathcal{O}| - 1\)}
  \For{each \(\ell = 1\) to \(L\) \textbf{in parallel}}
    \State Select machine \(m\) according to strategy \(\ell\)
    \State \(\mathcal{Q}^{(\ell)} \gets \emptyset\)
    \ForAll{\(Ps \in \mathcal{B}_t^{(\ell)}\)}
      \ForAll{\(o \in \mathrm{NextOps}(Ps),\ \text{where } M(o) = m\)}
        \State \(Ps' \gets Ps \cup \{o\}\)
        \State Evaluate score: \(\Phi^{(\ell)}(Ps') \gets C_{\max}(Ps') - \pi^{(\ell)}(M(o))\)
        \State \(\mathcal{Q}^{(\ell)} \gets \mathcal{Q}^{(\ell)} \cup \{Ps'\}\)
      \EndFor
    \EndFor
    \State \(\mathcal{B}_{t+1}^{(\ell)} \gets\) top(\(K\)) \(Ps\) in \(\mathcal{Q}^{(\ell)}\) by lowest \(\Phi^{(\ell)}\)
    \State With probability \(\epsilon\), randomly replace elements in \(\mathcal{B}_{t+1}^{(\ell)}\) from \(\mathcal{Q}^{(\ell)} \setminus \mathcal{B}_{t+1}^{(\ell)}\)
  \EndFor
\EndFor
\State \Return \(\arg\min_{Ps \in \cup_{\ell=1}^L \mathcal{B}_{|\mathcal{O}|}^{(\ell)}} C_{\max}(Ps)\)
\end{algorithmic}\end{minipage}
}
\end{algorithm}

Priority functions \(\pi^{(\ell)}\) steer each instance toward different machine usage profiles. Noise parameter \(\epsilon\) prevents premature convergence and encourages exploration. PMS-BS enhances solution robustness, particularly when used on multicore platforms.

The PMS-BS parallelization uses the master/slave paradigm: a master process maintains the global best makespan ${C_{\max}}^{\text{best}}$ and coordinates communication, while each slave process independently runs a beam search guided by a unique machine-priority strategy and randomization. The slaves periodically report their current best complete solution to the master, which updates $C_{\max}^{\text{best}}$ if an improvement is found. The master then broadcasts the updated value to all slaves, allowing them to prune partial schedules with estimated makespan exceeding $C_{\max}^{\text{best}}$, thus improving efficiency without requiring synchronization of internal search logic.

\section{GPU-Accelerated PMS-BS (G-PMS-BS)}
\label{sec:gpu}
The PMS-BS algorithm evaluates a large volume of candidates. Expanding a beam width of tens of thousands by up to $J$ operations requires simulating hundreds of thousands of partial schedules per iteration. Multi-core CPUs lack the parallel throughput to sustain this workload. 
G-PMS-BS addresses this bottleneck by mapping the beam search expansion and evaluation directly onto GPU hardware.
\subsection{GPU Architecture}
\label{subsec:gpu-background}
\begin{figure}[H]
    \centering
    \includegraphics[width=0.95\columnwidth]{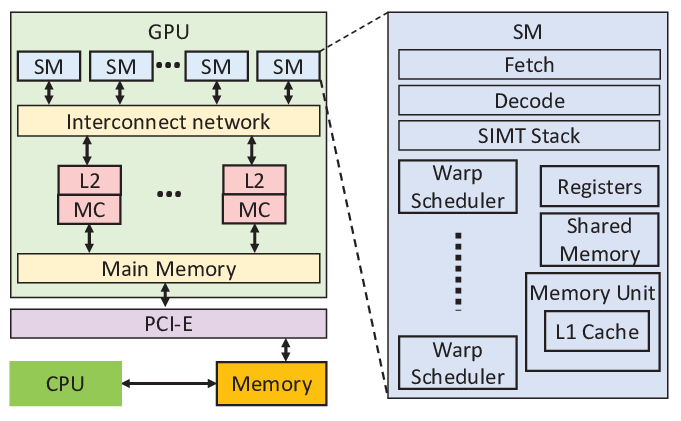}
    \caption{GPU architecture overview with an array of Streaming Multiprocessors (SMs) with hierarchical memory connected to the host CPU.}
    \label{fig:gpu_arch}
\end{figure}

Graphics Processing Units (GPUs) are designed for massively parallel computation capable of executing thousands of threads simultaneously. 
 As depicted in \autoref{fig:gpu_arch} in A GPU contains multiple Streaming Multiprocessors (SMs), and each SM has large number of processing unites. When we launch a GPU kernel (function), we define a \textit{grid} of \textit{thread blocks}; each block is assigned to an SM, and threads within the block are grouped into \textit{warps} of 32. The warp is the smallest scheduling unit: all threads in a warp execute the same instruction on different data (SIMD model).

 The GPU offers several levels of memory hierarchy, each with different latency and capacity:   \textit{Registers} are the fastest storage (1-2 cycles) private to each thread. However, the number of registers per GPU is limited, and using large number of registers limit the amount of parallelism you get from the GPU.  
\textit{shared memory} (2-5 cycles) offer low-latency read/write memory for threads within the same block, used to share data between threads. \textit{ Global memory} provides the largest capacity, but also the slowest with 400–800 clock cycles for uncached accesses.  This memory is accessible by all threads for read/write operations. 

To maximize parallelism, we balance these resources to minimize costly global memory traffic.

\subsection{G-PMS-BS and Mapping on GPUs}

\begin{figure*}
    \includegraphics[width=1.99\columnwidth]{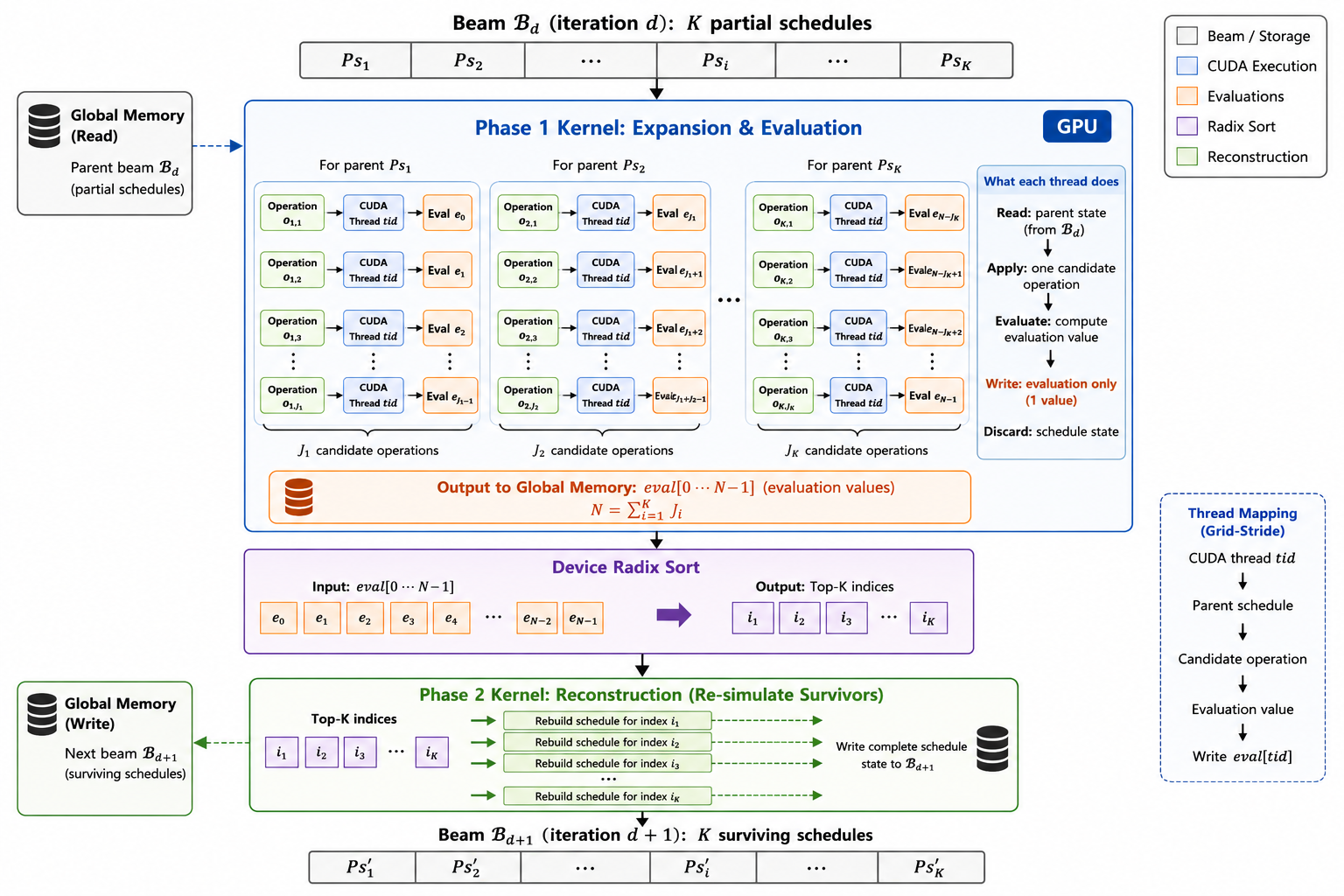}
    \caption{GPU thread-level mapping at iteration~\(d\). Each thread independently extends a parent partial schedule \(Ps_i\) by inserting one feasible operation \(o\) and writes the partial Cmax evaluation score. The device radix sort selects the top-\(K\) indices, and the Phase~2 kernel reconstructs the full state for the \(k\) partial schedules.}
\label{fig:gpu_thread_expansion}
\end{figure*}

The GPU-accelerated multi-strategy beam search (G-PMS-BS) maps the beam expansion and evaluation step onto the GPU by exposing fine‑grained parallelism while limiting intermediate memory usage. Storing generated child data during the beam expansion would consume most of the global capacity and create a memory bottleneck, preventing the algorithm to scale for large problem instances. As shown in ~\autoref{fig:gpu_thread_expansion}, each iteration of the beam search is split into three sequential stages: candidate expansion and evaluation, device‑side selection, and schedule reconstruction. This decomposition enables the GPU to evaluate many candidate schedules concurrently while only reconstructing the subset that survives selection.

\paragraph{Expansion and Evaluation.}
At iteration \(d\), the current beam \(\mathcal{B}_d\) contains \(K\) partial schedules. Each partial schedule generates a set of feasible next candidate operations. The union of all such candidates forms a one‑dimensional work list. G-PMS-BS uses a flattened grid‑stride mapping: each GPU thread independently evaluates one or more candidate operations. For each candidate, the thread simulates the required scheduling decision (including all relevant constraints), computes and stores the partial Cmax with a unique candidate id. As a result, we don't generate and store the schedule states at this stage avoiding huge memory footprint. 

\paragraph{Selection}
After all candidate scores have been computed, they are ranked entirely on the GPU using a device‑wide radix sort. The sorted list directly identifies the \(K\) globally best candidates across all parent schedules. Performing the ranking on the device avoids transferring the entire candidate set to the host and preserves the parallelism achieved during evaluation. 

\paragraph{Reconstruction}
The reconstruction kernel processes only the selected candidates. Each thread replays the scheduling decisions for one surviving candidate and reconstructs its complete scheduling state before writing it into the next beam  iteration\(\mathcal{B}_{d+1}\). By reconstructing only the survivors, G-PMS-BS reduces intermediate memory usage while maintaining the same search behavior as the original beam search.

\paragraph{Multi‑GPU execution}
To exploit multiple GPUs, G-PMS-BS adopts a distributed‑memory design based on MPI. Each MPI rank controls a single GPU and runs an independent beam‑search instance. Different machine‑priority strategies are assigned to different ranks to encourage search diversity. At configurable synchronization intervals, the ranks exchange the current global best Cmax value via collective communication. This allows all instances to prune inferior schedules while continuing their own search independently. The resulting design keeps communication overhead low and enables efficient scaling across multiple GPUs.
\paragraph{Profiling and Roofline}

We profiled the G-PMS-BS execution on the ta80 problem instance ($100 \times 20$) using NVIDIA Nsight Systems and Nsight Compute profiling tools. 

\begin{figure}[H]
    \centering
    \includegraphics[width=\columnwidth, trim=0 11cm 0 0, clip]{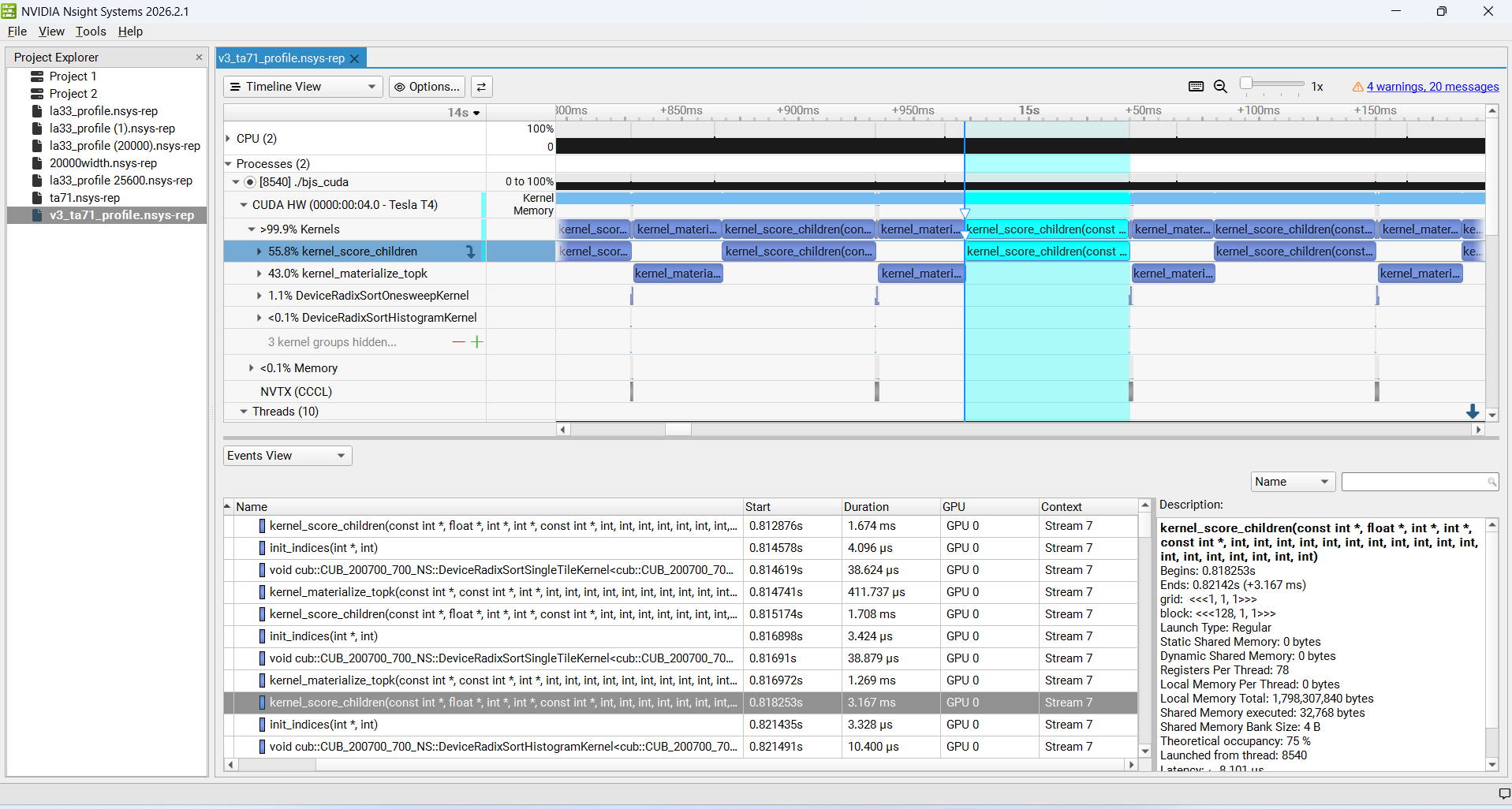}
    
    \includegraphics[width=1\columnwidth]{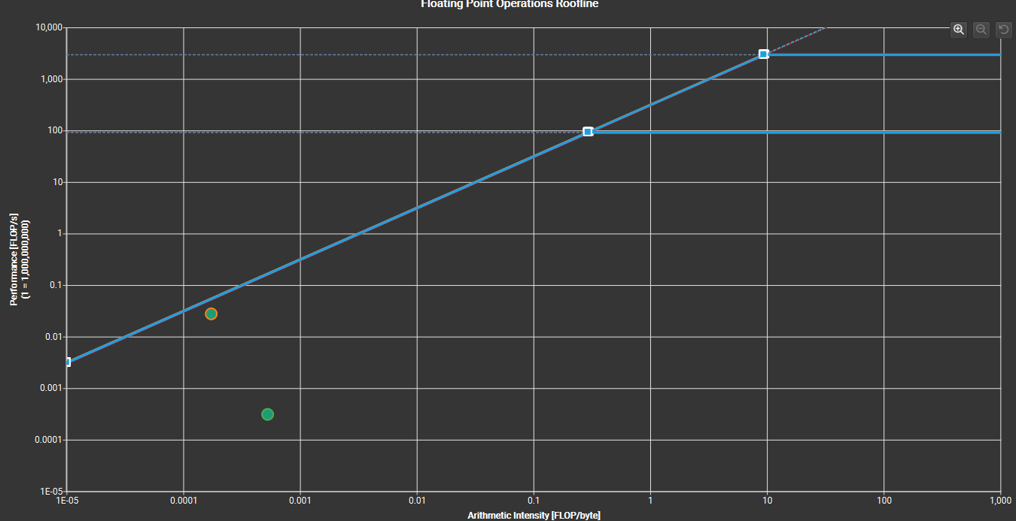}
    \caption{Nsight system (top) and Nsight compute (bottom) profiling for the GPU version.}
    \label{fig:gpu_execution}
\end{figure}

The sight Systems timeline in Figure~\ref{fig:gpu_execution} shows continuous execution without host-device communication bottlenecks. Phase 1 (Scoring) accounts for $55.8\%$ of the GPU time, Phase 2 (Reconstruction) takes $43.0\%$, and the parallel sort takes $1.1\%$.

As shown in \autoref{fig:gpu_execution}, The Roofline model places both kernels in the memory‑bound region which means that both kernels do more memory operations than arithmetic operations.  The reconstruction kernel is additionally latency bound due to the high thread divergence since each thread regenerates a different schedule state, leading to divergent control flow and scattered global memory accesses. 
Changing the mapping so that each block handles the regeneration of a contiguous group of surviving candidates would reduce divergence and improve coalescing, potentially alleviating this latency bottleneck.

\begin{figure*}[t!]
    \centering
    \begin{tikzpicture}
        \begin{axis}[
            xlabel={Beam size $K$},
            ylabel={Makespan $C_{\max}$},
            ymin=1150, ymax=1660,
            xmin=0, xmax=500,
            xtick={0,100,200,300,400,500},
            ytick={1200,1300,1400,1500,1600},
            legend style={at={(0.02,0.98)}, anchor=north west, font=\small, fill=white, draw=black},
            grid=major,
            width=1.8\columnwidth,
            height=0.9\columnwidth,
            axis y line*=left, 
        ]
        \addplot[color=blue, mark=*, mark size=2pt, thick] coordinates {
            (1,1304) (20,1235) (30,1198) (40,1193) (50,1191) (60,1198)
            (90,1180) (100,1180) (200,1180) (300,1174) (400,1168) (500,1160)
        };
        \addlegendentry{BS-ICH with machine selection}

        \addplot[color=black, mark=square*, mark size=2pt, thick] coordinates {
            (1,1553) (20,1318) (30,1278) (40,1553) (50,1494) (60,1511)
            (90,1351) (100,1436) (200,1312) (300,1287) (400,1403) (500,1411)
        };
        \addlegendentry{BS-ICH without machine selection}
    \end{axis}
    \begin{axis}[
            xlabel={Beam size $K$},
            ylabel={Time (seconds)},
            ymin=0, ymax=600,
            xmin=0, xmax=500,
            xtick=\empty, 
            ytick={0,100,200,300,400,500,600},
            legend style={at={(0.85,0.85)}, anchor=south east, font=\small, fill=white, draw=black},
            grid=major,
            width=1.8\columnwidth,
            height=0.9\columnwidth,
            axis y line*=right,
            axis x line=none,
        ]
        \addplot[color=red, mark=triangle*, mark size=2pt, thick] coordinates {
            (1,1.44) (20,22) (30,33) (40,43) (50,53) (60,73)
            (90,92) (100,100) (200,200) (300,300) (400,400) (500,500)
        };
        \addlegendentry{Computation time}
    \end{axis}
    \end{tikzpicture}
    \caption{BS-ICH makespan and execution time versus beam size $K$. Machine selection yields monotonic makespan improvement; computation time grows linearly.}
    \label{fig:66}
\end{figure*}
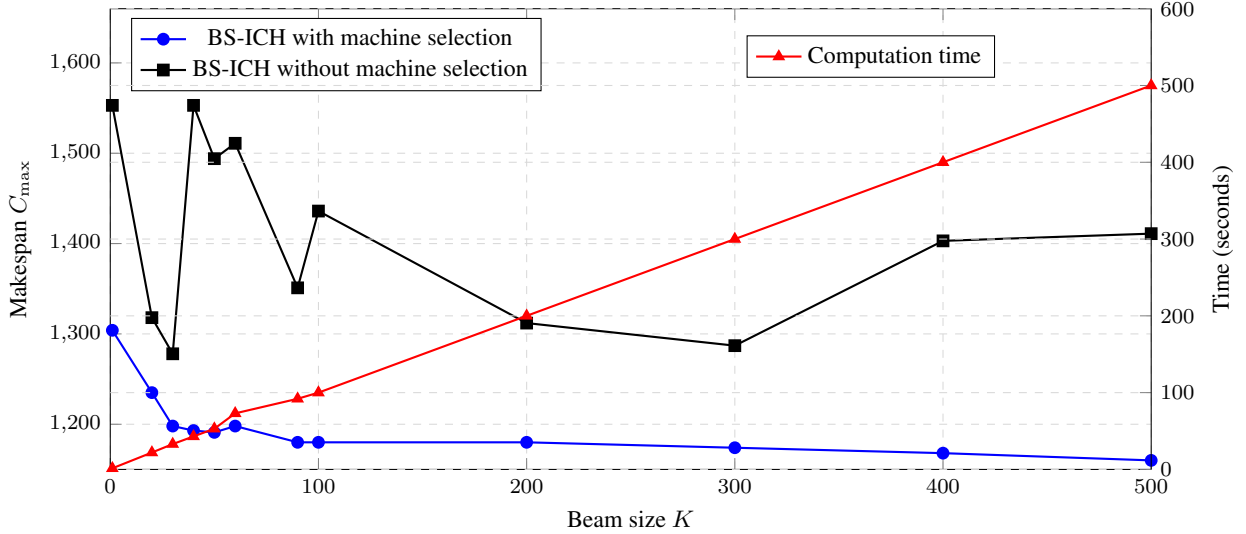
\subsection{Hybrid CPU+GPU Execution}

Although G-PMS-BS fully exploits the massive throughput of the GPU, a GPU-only {execution leaves} the host CPU cores {largely idle. The hybrid execution model reclaims this otherwise-wasted compute capacity by assigning the otherwise-idle MPI ranks to CPU-based beam searches. Specifically,  ranks \(0-3\) drive the GPU kernels with a wide beam of width \(K_{\text{GPU}}\). Ranks \(4\) through \(Nmpi-1\) (where \(Nmpi\) is the total number of MPI ranks) execute the CPU beam-search routine described in Section~VII, using a narrow beam of width \(K_{\text{CPU}}\). All ranks operate completely independently during the search loop: there is no inter-rank communication, and no global-best pruning occurs within the beam iterations. This design eliminates synchronization overhead entirely and ensures that GPU and CPU searches never block each other. At the end of the wall-clock budget, the global best makespan is gathered via a single \texttt{MPI\_Reduce} operation. 

\paragraph{Lower-Bound Scoring and Ranking}

{Search diversity is driven} as much by the scoring function as by the {initial} machine-priority seeds. {Rather than} ranking partial schedules {only by their partial makespan \(C_{\max}(P_s)\), each rank may adopt} one of several admissible lower bounds. {These bounds are maintained incrementally in \(O(1)\) time from three running data variables:} the remaining processing time on each machine (\(\mathtt{mach\_rem}[m]\)), the remaining processing time of each job (\(\mathtt{job\_rem}[j]\)), and the {ready} time of each job's last scheduled operation (\(\mathtt{job\_ready}[j]\)). The five scoring modes are defined as:

\begin{align}
\Phi_0 &= C_{\max} \quad \text{(baseline, LB\_CUR)}, \tag{1} \\
\Phi_1 &= \max\bigl(C_{\max},\; \max_m(\mathtt{mach\_ready}[m] + \mathtt{mach\_rem}[m])\bigr) \notag \\
&\qquad \text{(LB\_M, mode 1)}, \tag{2} \\
\Phi_2 &= \max\bigl(\Phi_1,\; \max_j(\mathtt{job\_ready}[j] + \mathtt{job\_rem}[j])\bigr) \notag \\
&\qquad \text{(LB\_MJ, mode 2)}, \tag{3} \\
\Phi_3 &= \max\Bigl(\Phi_1,\; \max_m \bigl(\min_{i\in U_m} h_i + \sum_{i\in U_m} p_i + \min_{i\in U_m} q_i\bigr)\Bigr) \notag \\
&\qquad \text{(LB\_HT, mode 3)}, \tag{4} \\
\Phi_4 &= \max\bigl(\Phi_1,\; \max_{i\in U}(h_i + p_i + q_i)\bigr) \quad \text{(LB\_CP, mode 4)}. \tag{5}
\end{align}

where \(U_m\) is the set of unscheduled operations on machine \(m\), \(U\) is the full unscheduled set, \(h_i = \max(\mathtt{job\_ready}[j_i], \mathtt{mach\_ready}[m_i])\) is the earliest feasible start time of operation \(i\), and \(q_i = \mathtt{tail\_static}[i]\) is the static suffix sum of processing times following operation \(i\) within its job, precomputed once. Rank \(r\) is statically assigned mode \(r \bmod 5\), ensuring that all five bounds contribute simultaneously without requiring explicit coordination. Tighter bounds (LB\_HT, LB\_CP) produce a smaller but higher-quality beam, while looser bounds preserve broader coverage of the search space. The global-best reduction at the end of each layer evaluates the best complete schedule across all five trajectories, combining the strengths of each bound.

\paragraph{CPU Beam-Width Calibration}

A GPU rank evaluates one to two orders of magnitude more partial schedules per second than a CPU rank, so without adjustment the CPU workers would lag behind and contribute little. Rather than relying on a fixed ratio or manual tuning, each CPU rank performs a short calibration run at startup. The calibration executes \(K_{\text{calib}}\) beam iterations (typically \(K_{\text{calib}} = 50\)) and measures the elapsed time. The CPU beam width is then scaled linearly to target a user-defined iteration time (typically \(0.5\) seconds):

\begin{equation}
K_{\text{CPU}} \gets \min\!\left( K_{\max},\; \max\!\left( K_{\min},\; K_{\text{calib}} \cdot \frac{T_{\text{target}}}{T_{\text{calib}}} \right) \right)
\end{equation}

where \(T_{\text{calib}}\) is the measured time for the calibration run and \(T_{\text{target}}\) is the desired iteration time. This automatic calibration eliminates manual tuning, adapts to the instance size and hardware, and ensures that CPU workers remain productive without falling too far behind the GPU ranks. If the user provides an explicit beam width via the command line, that value is used instead and the calibration is skipped. A worker's scoring mode is left unchanged by the adjustment, preserving the diversity of bounds across the hybrid ensemble.

\begin{figure}[H]
    \centering
    \includegraphics[width=1\linewidth]{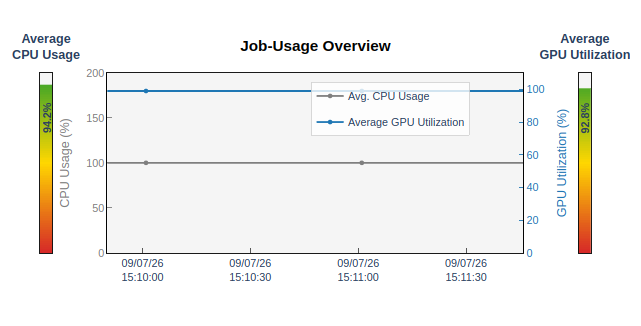}
    \caption{LLview monitoring report showing near-saturated utilization: 94\% CPU and 93\% GPU under the hybrid execution model.   }
    \label{fig:utilization2}
\end{figure}

\paragraph{Hardware Utilization}

Performance monitoring via LLView on the JURECA supercomputer quantifies the resource shift induced by the hybrid rank assignment. In GPU-only mode, the GPU is saturated at \(98.1\%\) utilization, while the host CPU remains largely idle at \(3.1\%\), serving only as a lightweight dispatcher. Under the hybrid mode, the CPU cores used by the CPU ranks reach \(94\%\) utilization (averaged across all 128 cores per node in core-level heatmaps), while GPU utilization experiences only a modest reduction to \(93\%\). This demonstrates that the CPU ranks successfully harvest idle host capacity without degrading GPU performance. The slight drop in GPU occupancy is an acceptable trade-off, as the CPU-discovered solutions contribute directly to the global best. Since all ranks search independently and only exchange the final best value, the search remains fully scalable and free of synchronization bottlenecks. ~\autoref{fig:utilization2} illustrates this complementary resource utilization for LA24 instances using one JURECA node with four A100 and 128 AMD CPU cores.

\section{Experimental Evaluation}
\label{sec:experiments}

In this section we assess the performance of all three proposed methods: BS-ICH, PMS-BS, and GPU-accelerated G-PMS-BS against state-of-the-art approaches using Lawrence~\cite{15} and Taillard~\cite{taillard1993benchmarks} benchmarks under the BJSSP with swap allowed.

All experiments run on the JURECA-DC supercomputer~\cite{JUWELS} at J\"ulich Supercomputing Centre. Each node has two AMD EPYC 7742 processors (128 cores) and four NVIDIA A100 GPUs (40\,GB HBM2). The code is written in C++/CUDA with MPI and OpenMP.

Figure~\ref{fig:66} shows the BS-ICH makespan variation when increasing the size of the beam with and without machine selection. It illustrates the relationship between the makespan and the beam size $k$, alongside the computational time required. The results demonstrate that BS-ICH with machine selection at each iteration achieves a significant reduction in makespan as the beam size $k$ increases. This improvement is attributed to the more extensive exploration of the solution space. In contrast, BS-ICH without machine selection shows limited makespan improvement and irregular behavior despite increasing beam size; this is due to the beam members converging on nearly identical partial schedules, resulting in redundant search paths. The results underline the critical importance of machine selection in maintaining diversity within the beam and effectively navigating the solution space to achieve better makespan. Additionally, the computational time grows linearly with the beam size, as shown by the red curve.

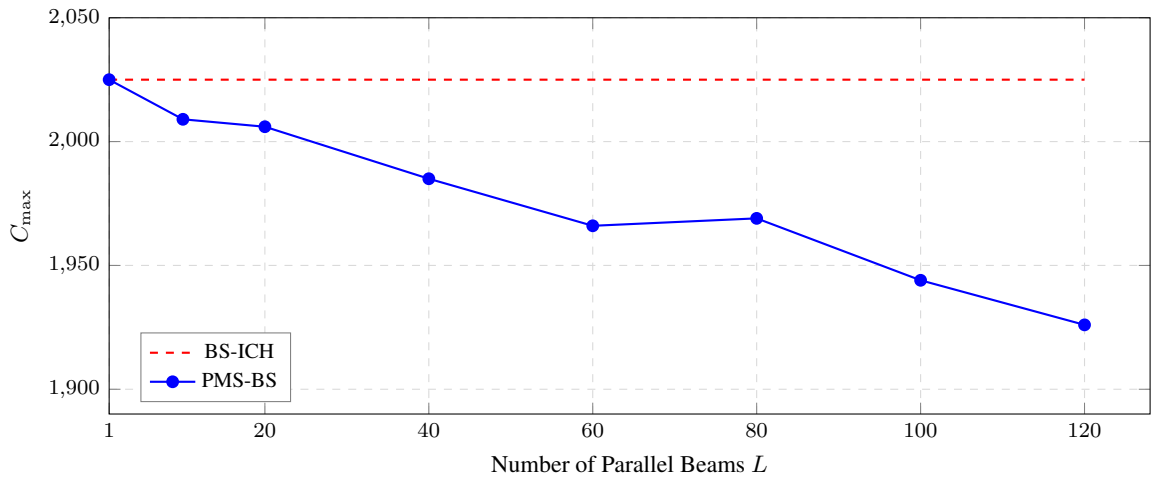
\begin{figure*}[h!]
    \centering
    \begin{tikzpicture}
      \begin{axis}[
        xlabel={Number of Parallel Beams $L$},
        ylabel={$C_{\max}$},
        ymin=1890, ymax=2050,
        xmin=1, xmax=128,
        xtick={1,20,40,60,80,100,120},
        ytick={1900,1950,2000,2050},
        legend pos=south west,
        grid=both,
        width=1.8\columnwidth,
        height=0.8\columnwidth,
      ]
        \addplot[red, thick, dashed] coordinates {(1,2025)  (120,2025)};
        \addlegendentry{BS-ICH}
        \addplot[blue, thick, mark=*, mark size=2pt] coordinates {
          (1,2025) (10,2009) (20,2006) (40,1985) (60,1966) (80,1969) (100,1944) (120,1926)
        };
        \addlegendentry{PMS-BS}
      \end{axis}
    \end{tikzpicture}
    \caption{Impact of increasing the number of parallel beams $L$ on the makespan result for LA30 instance.}
    \label{fig:l}
\end{figure*}

Figure~\ref{fig:l} compares BS-ICH and PMS-BS as the number of parallel beams $L$ increases for La30 instance. BS-ICH has a single beam with a flat makespan, while PMS-BS keeps improving the makespan. 
Indeed, since BS-ICH selects a single machine with each iteration, it easily becomes trapped in a local optimum. PSM-BS, on the other hand, has multiple parallel instances that diversify the search. Each instance prioritizes a distinct machine. As a result, local optima are avoided and the explored search space increases, leading to better solution quality. 

Note that the goal of this parallelization is to explore more search space, not to reduce execution time, which remains similar. 
This motivates our GPU-accelerated extension, where hundreds of strategies can run simultaneously on a GPU (Faster runtime and more solution space exploration). 

\renewcommand{\arraystretch}{1}
\begin{table*}[h!]
\centering
\scriptsize
\caption{Comparison of makespan results for Lawrence instances LA01--LA40. Lower is better. Underlined values are the best across all methods. Results using 4xA100 with 128 CPU cores.}
\label{tab:la_comparison}
\begin{tabular}{|c|c|c|c|c|c|c|c|}
\hline
\textbf{Instance} & \textbf{PTS-MPSS}~\cite{dabah2019efficient} &
\textbf{IFS/CP-OPT}~\cite{9} & \textbf{Par.\ B\&B}~\cite{d3} &
\textbf{BS-ICH} & \textbf{PMS-BS} & \textbf{G-PMS-BS} & \textbf{H-PMS-BS} \\
\hline
LA01 & \underline{793} & \underline{793} & \underline{793} & \underline{793} & \underline{793} & \underline{793} & \underline{793} \\
LA02 & \underline{793} & \underline{793} & \underline{793} & \underline{793} & \underline{793} & \underline{793} & \underline{793} \\
LA03 & \underline{715} & \underline{715} & \underline{715} & 747 & \underline{715} & \underline{715} & \underline{715} \\
LA04 & \underline{743} & \underline{743} & \underline{743} & 776 & \underline{743} & \underline{743} & \underline{743} \\
LA05 & \underline{664} & 686 & \underline{664} & 680 & \underline{664} & \underline{664} & \underline{664} \\
LA06 & 1092 & 1064 & \underline{1060} & 1142 & 1062 & 1081 & \underline{1060} \\
LA07 & 1026 & 1038 & \underline{1016} & 1061 & 1020 & 1022 & 1020 \\
LA08 & 1060 & 1062 & \underline{1040} & 1105 & \underline{1040} & \underline{1040} & \underline{1040} \\
LA09 & 1183 & 1185 & \underline{1141} & 1206 & \underline{1141} & \underline{1141} & \underline{1141} \\
LA10 & 1110 & 1110 & \underline{1096} & 1165 & 1121 & 1123 & 1111 \\
LA11 & 1445 & 1466 & 1465 & 1519 & 1417 & 1424 & \underline{1404} \\
LA12 & 1269 & 1272 & 1253 & 1323 & 1245 & 1234 & \underline{1224} \\
LA13 & 1440 & 1465 & 1383 & 1456 & 1382 & \underline{1375} & 1383 \\
LA14 & 1465 & 1548 & 1467 & 1547 & 1451 & 1450 & \underline{1429} \\
LA15 & 1515 & 1527 & 1511 & 1561 & \underline{1453} & 1459 & 1455 \\
LA16 & \underline{1060} & 1084 & \underline{1060} & 1110 & \underline{1060} & \underline{1060} & \underline{1060} \\
LA17 & 930 & 930 & \underline{929} & 964 & \underline{929} & \underline{929} & \underline{929} \\
LA18 & \underline{1025} & 1026 & \underline{1025} & 1080 & 1026 & 1026 & 1026 \\
LA19 & \underline{1043} & \underline{1043} & \underline{1043} & 1088 & \underline{1043} & 1059 & \underline{1043} \\
LA20 & 1074 & 1074 & \underline{1060} & 1109 & 1080 & 1080 & 1069 \\
LA21 & 1410 & 1521 & 1433 & 1493 & 1418 & 1421 & \underline{1408} \\
LA22 & 1303 & 1425 & 1424 & 1385 & 1303 & 1330 & \underline{1302} \\
LA23 & \underline{1338} & 1531 & 1536 & 1530 & 1433 & 1431 & 1393 \\
LA24 & 1399 & 1498 & 1354 & 1486 & 1359 & 1359 & \underline{1348} \\
LA25 & 1343 & 1424 & \underline{1313} & 1469 & 1364 & 1353 & 1342 \\
LA26 & 1914 & 2045 & 1994 & 1958 & 1830 & 1810 & \underline{1782} \\
LA27 & 1959 & 2104 & 2005 & 2053 & 1888 & \underline{1854} & 1862 \\
LA28 & 1878 & 2027 & 2020 & 2053 & 1893 & 1823 & \underline{1798} \\
LA29 & 1819 & 1963 & 1725 & 1893 & 1735 & 1708 & \underline{1685} \\
LA30 & 1942 & 2095 & 2055 & 2008 & 1860 & \underline{1839} & 1858 \\
LA31 & 2748 & 3078 & 3028 & 2927 & 2769 & \underline{2526} & 2554 \\
LA32 & 2935 & 3336 & 3368 & 3133 & 2914 & \underline{2742} & 2763 \\
LA33 & 2662 & 3147 & 3037 & 2884 & 2620 & 2498 & \underline{2492} \\
LA34 & 2731 & 3125 & 2961 & 2956 & 2679 & \underline{2550} & 2577 \\
LA35 & 2722 & 3148 & 3065 & 2991 & 2742 & \underline{2586} & 2601 \\
LA36 & 1685 & 1793 & 1701 & 1862 & 1706 & 1696 & \underline{1665} \\
LA37 & 1815 & 1983 & 1848 & 1981 & 1784 & \underline{1773} & 1788 \\
LA38 & 1606 & 1708 & 1598 & 1771 & 1605 & 1591 & \underline{1576} \\
LA39 & 1705 & 1783 & 1714 & 1776 & 1681 & 1666 & \underline{1664} \\
LA40 & 1708 & 1777 & 1714 & 1805 & 1629 & \underline{1617} & 1630 \\
\hline
\end{tabular}
\end{table*}

Table~\ref{tab:la_comparison} presents a comparison of all four proposed configurations against three state-of-the-art methods holding the best known solutions to our knowledge, across the complete Lawrence benchmark suite (LA01--LA40). The results reveal a clear progression: each method in our framework addresses the limitations of its predecessor, and the relative advantage grows with instance size.

On the optimally solvable instances (LA01--LA10, LA16--LA20), Parallel B\&B~\cite{d3} holds the optimal results, and both PMS-BS and G-PMS-BS match them in 11/15 instances. BS-ICH, on the other side, gets stuck in local optima due to its fixed machine-selection order, while PMS-BS addresses this through machine-priority biases and parallel diversification.
On the remaining~4 instances, the gaps are very small: LA07 ($+0.4\%$), LA10 ($+1.36\%$), LA18 ($+0.1\%$), and LA20 ($+0.84\%$), yielding an average optimality gap of only~$0.67\%$ across all four instances. It is worth mentioning despite our approaches are  a pure construction heuristics with no improvement phase, we achieve near-optimal solutions on instances with proven lower bounds.

On medium and large instances (LA11--LA15, LA21--LA40), both parallel B\&B and PTS retain best results for LA23 and LA25 respectively. For all remaining 23 instances our proposed GPU and hybrid (G-PMS-BS and H-PMS-BS) approaches achieve new best known results. The improvements are particularly significant, on the 30$\times$10 subset for example, LA31 drops from 2748 (PTS) to 2526 ($-$8\%) and LA33 from 2662 (PTS) to 2492 ($-$7\%). This significant improvement is the result of the following:
(i) the new construction strategy that retains the $k$ best schedules,
(ii) the diversification strategies that escape local optima, and
(iii) the large number of solutions explored thanks to the processing power of GPUs.

Table~\ref{tab:la_comparison} shows a clear progression in solution quality, where each proposed method addresses the main limitation of its predecessor. {BS-ICH} constructs only feasible schedules by maintaining the $K$ best partial solutions, but its deterministic machine-selection strategy causes the search to converge in local optima. {PMS-BS} alleviates this limitation by launching multiple independent beam searches with different machine-priority on multi-core CPUs. This diversification enables much better solutions than BS-ICH.

{G-PMS-BS} further improves PMS-BS by exploiting the massive parallelism of GPUs. Thousands of candidate schedules are evaluated concurrently, enabling beam widths of up to 25,600 and a much broader exploration of the search space within the same time budget. The larger beam significantly reduces the risk of pruning promising partial schedules, leading to superior solution quality, particularly on large and difficult instances.

Finally, {H-PMS-BS} combines large-scale GPU exploration with lower-bounds that estimate the remaining scheduling cost, allowing more informed beam ranking and pruning. Consequently, the hybrid approach not only explores more candidate schedules but also explores them more intelligently, producing the best overall makespan results across the benchmark instances.

\begin{table}[h!]
\centering
\caption{Multi‑node scaling: 4 GPUs (1 node) vs. 16 GPUs (4 nodes) for LA31–LA36. All runs used a 10‑minute wall‑clock limit.}
\label{tab:scaling_large}
\begin{tabular}{lcc}
\toprule
\textbf{Instance} & \textbf{4 GPUs (\(C_{\max}\))} & \textbf{16 GPUs (\(C_{\max}\))} \\
\midrule
LA31 & 2526 & 2532 \\
LA32 & 2742 & \textbf{2705} \\
LA33 & 2498 & \textbf{2466} \\
LA34 & 2550 & \textbf{2521} \\
LA35 & 2586 & \textbf{2574} \\
LA36 & 1696 & \textbf{1666} \\
\bottomrule
\end{tabular}
\end{table}

Table~\ref{tab:scaling_large} presents the best makespans obtained for the largest Lawrence instances (LA31–LA36) using 4 GPUs (single node) and 16 GPUs (four nodes) with the same wall‑clock time (10 minutes). Increasing the number of GPUs consistently improves solution quality, with relative reductions ranging from 0.3\% (LA31) to 1.4\% (LA32). The improvement is most pronounced on LA32 and LA33, where the additional parallelism enables deeper exploration of the search space and escape from local optima. These results demonstrate that the proposed GPU‑accelerated beam search scales positively with more hardware, providing better makespans without increasing runtime.
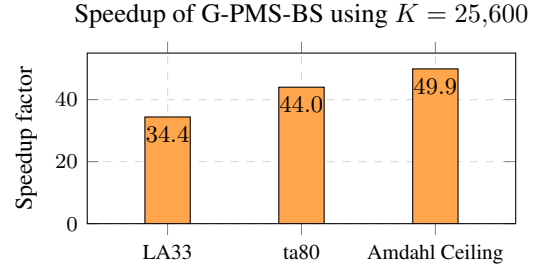
\begin{figure}[htbp]
    \centering
    \begin{tikzpicture}
        \begin{axis}[
            ybar,
            bar width=0.6cm,
            enlarge x limits=0.3,
            symbolic x coords={LA33, ta80, Amdahl Ceiling},
            xtick={LA33, ta80, Amdahl Ceiling},
            ylabel={Speedup factor},
            ymin=0,
            ymax=55,
            grid=major,
            nodes near coords,
            nodes near coords align={vertical},
            nodes near coords style={
                font=\small\bfseries,
                /pgf/number format/fixed,
                /pgf/number format/fixed zerofill,
                /pgf/number format/precision=1,
                anchor=north,
            },
            tick label style={font=\footnotesize},
            ylabel style={font=\small},
            title={Speedup of G-PMS-BS using $K=25{,}600$},
        ]
        \addplot[fill=orange!70] coordinates {(LA33, 34.4) (ta80, 44)(Amdahl Ceiling, 49.9)};
        \end{axis}
    \end{tikzpicture}
    \caption{GPU speedup against the Amdahl ceiling. The implementation achieves $34.4\times$ on LA33 (300 operations) and $44.0\times$ on ta80 (2,000 operations).}
    \label{fig:speedup}
\end{figure}

Figure~\ref{fig:speedup} reports the achieved speedup against the theoretical Amdahl ceiling for both a moderate LA33 and a large ta80 instances. For LA33 (300 operations), the GPU version achieves a $34.4\times$ speedup over the CPU baseline, cutting execution time from 4.1 hours to just 7.1 minutes, reaching 69\% of the Amdahl ceiling ($49.9\times$). The remaining gap is due to synchronization overhead in retrieving the K-best candidates and updating device data between iterations. On the much larger ta80 instance, the speedup rises to $44.0\times$, reaching 88\% of the Amdahl ceiling. This improvement comes from the higher GPU occupancy: the larger problem size provides substantially more parallel work per iteration to fully saturate the streaming multiprocessors and reduce kernel launch and synchronization overheads.

\subsection{Taillard Benchmark Results}

To evaluate extreme scalability, we run G-PMS-BS on all 80 Taillard instances, from 15$\times$15 (225 operations) to 100$\times$20 (2{,}000 operations). We compare against J.~K.~Mogali~(2021)~\cite{mogali2021}, who established the first comprehensive baseline for the Taillard dataset under blocking constraints using a tuned Tabu Search with a 30-minute execution limit per instance.  To handle these large instances, we ran the GPU version (G-PMS-BS) on 16 GPUs.  

\begin{table}[t]
\centering
\caption{G-PMS-BS vs.\ Mogali~\cite{mogali2021} on Taillard TA01--TA80 ($K\!=\!25{,}600$, 16$\times$A100). Underlined values are the best of ten iterations.}
\label{tab:taillard_sota}
\scriptsize
\setlength{\tabcolsep}{2pt}
\renewcommand{\arraystretch}{0.88}
\begin{tabular}{@{}lcrrr|lcrrr@{}}
\toprule
Inst. & $J{\times}M$ & \cite{mogali2021} & G-PMS-BS & Time &
{Inst.} & $J{\times}M$ & \cite{mogali2021} &G-PMS-BS & Time \\
\midrule
ta01 & 15$\times$15 & 1705 & \underline{1692} & 28s & ta41 & 30$\times$20 & 3516 & \underline{3388} & 216s \\
ta02 & 15$\times$15 & 1665 & \underline{1642} & 27s & ta42 & 30$\times$20 & 3474 & \underline{3253} & 218s \\
ta03 & 15$\times$15 & 1655 & \underline{1616} & 27s & ta43 & 30$\times$20 & 3345 & \underline{3227} & 224s \\
ta04 & 15$\times$15 & 1617 & \underline{1588} & 28s & ta44 & 30$\times$20 & 3516 & \underline{3259} & 214s \\
ta05 & 15$\times$15 & 1679 & \underline{1616} & 29s & ta45 & 30$\times$20 & 3472 & \underline{3289} & 226s \\
ta06 & 15$\times$15 & 1691 & \underline{1668} & 29s & ta46 & 30$\times$20 & 3444 & \underline{3283} & 215s \\
ta07 & 15$\times$15 & 1707 & \underline{1626} & 28s & ta47 & 30$\times$20 & 3396 & \underline{3259} & 229s \\
ta08 & 15$\times$15 & 1701 & \underline{1639} & 29s & ta48 & 30$\times$20 & 3427 & \underline{3255} & 228s \\
ta09 & 15$\times$15 & 1731 & \underline{1729} & 29s & ta49 & 30$\times$20 & 3394 & \underline{3217} & 227s \\
ta10 & 15$\times$15 & 1687 & \underline{1660} & 29s & ta50 & 30$\times$20 & 3514 & \underline{3301} & 181s \\
\addlinespace
ta11 & 20$\times$15 & \underline{2009} & 2017 & 53s & ta51 & 50$\times$15 & 5079 & \underline{4748} & 267s \\
ta12 & 20$\times$15 & 2156 & \underline{2135} & 51s & ta52 & 50$\times$15 & 5148 & \underline{4768} & 269s \\
ta13 & 20$\times$15 & \underline{2010} & 2011 & 53s & ta53 & 50$\times$15 & 5041 & \underline{4758} & 277s \\
ta14 & 20$\times$15 & 2070 & \underline{1979} & 52s & ta54 & 50$\times$15 & 5074 & \underline{4659} & 282s \\
ta15 & 20$\times$15 & 2031 & \underline{1987} & 55s & ta55 & 50$\times$15 & 4972 & \underline{4735} & 284s \\
ta16 & 20$\times$15 & 2168 & \underline{2111} & 54s & ta56 & 50$\times$15 & 5113 & \underline{4793} & 296s \\
ta17 & 20$\times$15 & 2229 & \underline{2145} & 52s & ta57 & 50$\times$15 & 5187 & \underline{4867} & 294s \\
ta18 & 20$\times$15 & 2125 & \underline{2070} & 53s & ta58 & 50$\times$15 & 5128 & \underline{4954} & 298s \\
ta19 & 20$\times$15 & 2053 & \underline{2026} & 53s & ta59 & 50$\times$15 & 4974 & \underline{4722} & 297s \\
ta20 & 20$\times$15 & 2125 & \underline{2021} & 53s & ta60 & 50$\times$15 & 5119 & \underline{4803} & 304s \\
\addlinespace
ta21 & 20$\times$20 & 2439 & \underline{2395} & 91s & ta61 & 50$\times$20 & 5756 & \underline{5418} & 514s \\
ta22 & 20$\times$20 & 2402 & \underline{2367} & 95s & ta62 & 50$\times$20 & 5890 & \underline{5528} & 501s \\
ta23 & 20$\times$20 & 2348 & \underline{2328} & 94s & ta63 & 50$\times$20 & 5530 & \underline{5249} & 517s \\
ta24 & 20$\times$20 & \underline{2431} & \underline{2431} & 98s & ta64 & 50$\times$20 & 5523 & \underline{5145} & 527s \\
ta25 & 20$\times$20 & 2329 & \underline{2296} & 95s & ta65 & 50$\times$20 & 5558 & \underline{5312} & 509s \\
ta26 & 20$\times$20 & 2467 & \underline{2377} & 97s & ta66 & 50$\times$20 & 5723 & \underline{5343} & 526s \\
ta27 & 20$\times$20 & 2529 & \underline{2526} & 98s & ta67 & 50$\times$20 & 5715 & \underline{5277} & 534s \\
ta28 & 20$\times$20 & 2434 & \underline{2305} & 97s & ta68 & 50$\times$20 & 5740 & \underline{5350} & 532s \\
ta29 & 20$\times$20 & 2489 & \underline{2446} & 99s & ta69 & 50$\times$20 & 5810 & \underline{5429} & 532s \\
ta30 & 20$\times$20 & \underline{2374} & 2378 & 97s & ta70 & 50$\times$20 & 5799 & \underline{5455} & 521s \\
\addlinespace
ta31 & 30$\times$15 & 3123 & \underline{2975} & 122s & ta71 & 100$\times$20 & 12153 & \underline{10563} & 1816s \\
ta32 & 30$\times$15 & 3159 & \underline{3025} & 123s & ta72 & 100$\times$20 & 11444 & \underline{10345} & 1813s \\
ta33 & 30$\times$15 & 3234 & \underline{3117} & 120s & ta73 & 100$\times$20 & 11766 & \underline{10637} & 1780s \\
ta34 & 30$\times$15 & 3221 & \underline{3064} & 124s & ta74 & 100$\times$20 & 11897 & \underline{10454} & 1881s \\
ta35 & 30$\times$15 & 3106 & \underline{2955} & 131s & ta75 & 100$\times$20 & 11476 & \underline{10386} & 1852s \\
ta36 & 30$\times$15 & 3146 & \underline{3034} & 124s & ta76 & 100$\times$20 & 12067 & \underline{10571} & 1833s \\
ta37 & 30$\times$15 & 3223 & \underline{3115} & 127s & ta77 & 100$\times$20 & 12265 & \underline{10694} & 1854s \\
ta38 & 30$\times$15 & 3071 & \underline{2941} & 127s & ta78 & 100$\times$20 & 11697 & \underline{10478} & 1773s \\
ta39 & 30$\times$15 & 3002 & \underline{2846} & 128s & ta79 & 100$\times$20 & 11870 & \underline{10555} & 1747s \\
ta40 & 30$\times$15 & 3026 & \underline{2928} & 130s & ta80 & 100$\times$20 & 11405 & \underline{10258} & 1735s \\
\bottomrule
\end{tabular}
\end{table}

Table~\ref{tab:taillard_sota} compares G-PMS-BS with the state-of-the-art Tabu Search of Mogali~\cite{mogali2021} on the 80 Taillard benchmark instances. G-PMS-BS achieves better makespans on 77 out of 80 instances, matches one instance (ta24), and is marginally inferior on only two instances (ta11 and ta13) by 8 and 1 makespan units, respectively. The performance advantage becomes increasingly pronounced as the problem size grows, with average makespan reductions of 5.5\% on the 50$\times$15 instances and 5.9\% on the 50$\times$20 instances.

The most significant gains are observed on the largest 100$\times$20 instances (2,000 operations), where the search space becomes extremely large. In these instances, the scalability of G-PMS-BS enables the exploration of substantially more candidate schedules through beam widths of up to 25,600 while maintaining practical execution times. This broader exploration significantly increases the probability of retaining high-quality partial schedules throughout the search, leading to makespan reductions of 10--13\% compared with the previous state of the art. For example, \texttt{ta71} improves from 12,153 to 10,563 ($-13.1\%$), while \texttt{ta77} decreases from 12,265 to 10,694 ($-12.8\%$). These results demonstrate that the proposed GPU beam search scales effectively with problem size, with its advantage increasing as the complexity of the scheduling problem grows.



\section{Conclusion}
\label{sec:conclusion}

This paper addressed the Blocking Job-Shop Scheduling Problem with three complementary heuristics that progressively scale from single-core to massively parallel GPU execution. The key insight across all three methods is the same: by maintaining a beam of K-best feasible partial schedules, we eliminate the costly repair phases that burden metaheuristics, while the machine-serialized construction preserves beam diversity. BS-ICH establishes this foundation. PMS-BS escapes local optima via parallel diversification. G-PMS-BS maps the beam expansion onto GPUs using a two-phase kernel that reduces memory footprint from gigabytes to megabytes, enabling large beam widths on 2,000-operation instances. A hybrid CPU+GPU mode further harvests idle host cores with lower-bound scoring diversity.

Experiments on all 40 Lawrence and 80 Taillard benchmarks validate the proposed approaches. GPU approaches set new best-known results on 18/20 large Lawrence instances and  improves 77 of 80 Taillard instances (with up to 13\% on 100×20 problems), and delivers a 44× GPU speedup. 

By combining guaranteed feasibility, beam diversity, and massive parallelism, we prove that construction-based search is a powerful paradigm for large-scale combinatorial scheduling.



Future research directions include adapting the proposed methods for classical JSSP and Flexible JSSP.

\section*{Data and Code Availability}
The Lawrence~\cite{15} and Taillard~\cite{taillard1993benchmarks} benchmark datasets are publicly available. Source code for the GPU-accelerated PMS-BS implementation and all experimental scripts are available from the corresponding author upon reasonable request.

\bibliographystyle{ieeetr}
\bibliography{ref}
\EOD

\end{document}